\documentclass[english,jkps,reprint,fleqn,showpacs,showkeys]{revtex4-1}
\usepackage[T1]{fontenc}
\usepackage[latin9]{inputenc}
\usepackage{verbatim}
\usepackage{amsmath}
\usepackage{amssymb}
\usepackage{graphicx}

\makeatletter
\usepackage{bm}

\usepackage[colorlinks,urlcolor=blue]{hyperref}
\usepackage[hyphenbreaks]{breakurl}
\usepackage{upgreek}

\makeatother

\usepackage{babel}
\begin{document}
\title[]{Three dimensional fast tracker for central drift chamber based level 1 trigger system in the Belle II experiment}

\author{E. \surname{Won}}

\author{J. B. \surname{Kim}}
\email{jbkim@hep.korea.ac.kr}

\thanks{Preprint submitted to Journal of the Korean Physical Society}

\affiliation{Department of Physics, Korea University, Seoul 02841}

\author{B. R. \surname{Ko}}

\affiliation{Center for Axion and Precision Physics Research, Institute of Basic
Science, Deajon 34051}


\date[]{Received 6 November 2017}
\begin{abstract}
The Belle II detector at the SuperKEKB accelerator has a level 1 trigger
implemented in field-programmable gate arrays. Due to the high luminosity
of the beam, a trigger that effectively rejects beam induced background
is required. A three dimensional tracking algorithm for the level
1 trigger that uses the Belle II central drift chamber detector response
is being developed to reduce the recorded beam background while having
a high efficiency for physics of interest. In this paper, we describe
the three dimensional track trigger that finds and fits track parameters
which we developed. 
\end{abstract}

\maketitle

\section{INTRODUCTION}

The Belle II detector \cite{BelleIITDR} is located at the asymmetric-energy
electron positron collider SuperKEKB \cite{SuperKEKB}. The high luminosity
of SuperKEKB will open the possibility to a new panorama of measurements
in heavy flavor physics \cite{BelleIITDR}. In order to take advantage
of the high luminosity, the trigger is required to have a high efficiency
for physics of interest while suppressing other physics processes
such as the beam-wall background that are not originated near the
interaction point (IP). Also due to the hardware limits of the vertex
detectors, the level 1 trigger is required to provide a trigger within
5 $\upmu$s . Due to these requirements, the level 1 trigger is being
developed with boards which have large field-programmable gate arrays
(FPGAs) and high speed optical transceivers \cite{level1triggerBelleII}. 

The level 1 trigger will utilize the sub-detector responses where
the main sub-detectors are the central drift chamber (CDC) \cite{CDC}
and electric calorimeter. The CDC will provide information for charged
particles. It consists of nine super-layers (SLs) of wires where five
of them have wires arranged parallel to the beam axis (axial) and
four of them have finite shift in the phi direction at the end plates
(stereo). The CDC readout electronic system will send information
to the level 1 trigger by optical transmission. By utilizing this
information, the level 1 trigger will be able to reconstruct the trajectories
of the charged particles in three dimensions. 

The level 1 CDC trigger \cite{CDCTrigger} is made up of six modules
which are the merger, track segment finder (TSF), event time finder
(ETF), two dimensional finder (2DF), neural network tracker (NNT)
\cite{NNTrigger}, and three dimensional tracker (3DT) . The merger
boards combine the information from the CDC readout boards and send
the information to the TSF boards. The TSF finds parts of the particle
trajectory called track segments (TS) using pattern matching. The
TS information is then sent to the rest of the system which are the
ETF, 2DF, NNT, and 3DT. The ETF determines the event timing which
is sent to the NNT and 3DT. The 2DF finds the two dimensional track
using Hough transformation \cite{hough} and sends the results to
the NNT and 3DT. The NNT uses a feed-forward neural network to determine
the track parameters and sends the information to the later stage
of the level 1 trigger. The 3DT does geometric transformations and
applies linear regressions to determine the track parameters which
are sign of the electric charge, transverse momentum ($p_{T}$), incident
angle ($\phi_{i}$), polar angle ($\theta$), and distance of track
origin from the IP respect to the $z$ axis ($z_{0}$). These information
are sent to the later stages of the level 1 trigger. 

In this paper, the 3DT we developed will be described. The algorithms
of four modules in the 3DT will be explained which are the TS map
maker, stereo TS finder, and $z_{0}$ fitter. 

\section{Design}

\begin{figure*}
\begin{centering}
\includegraphics[width=1\textwidth]{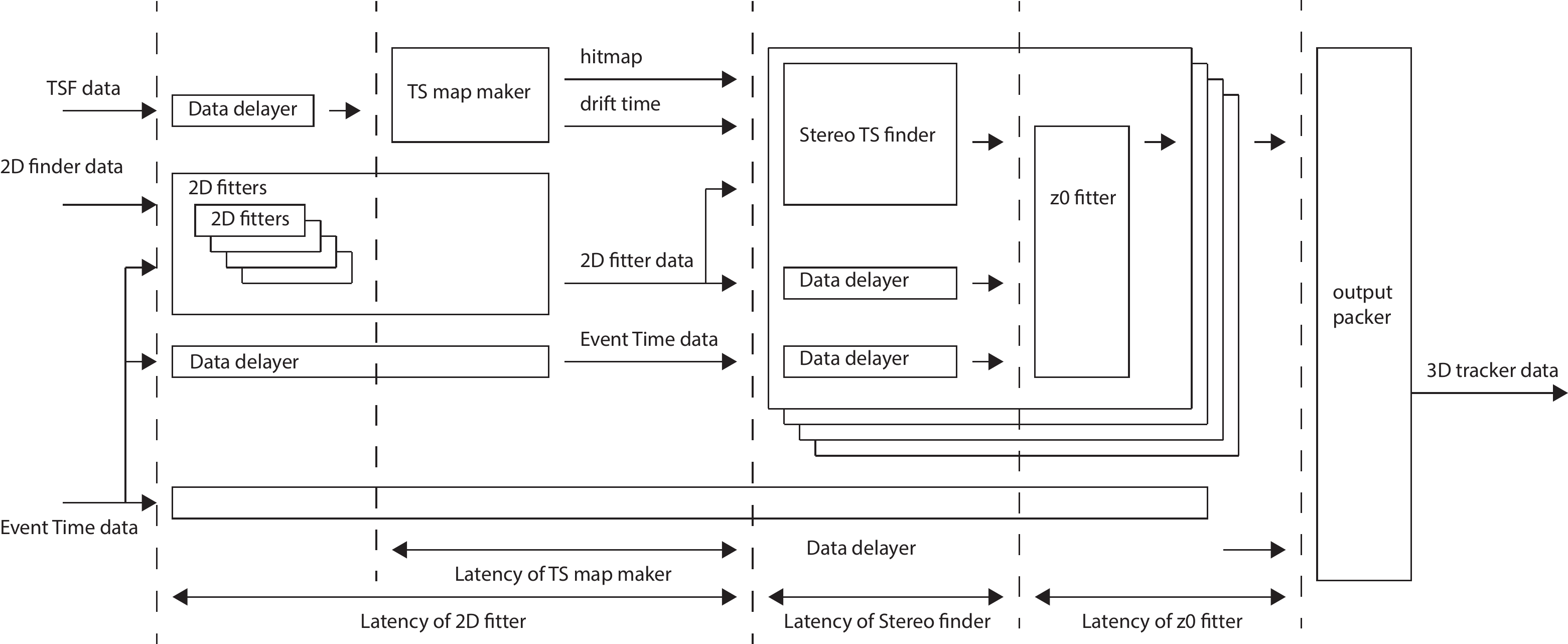}
\par\end{centering}
\caption{Structure of the 3DT. The diagram shows all the modules in the 3DT.
The connections and latency relationship between the modules are also
shown. The 3DT modules are data delayers, 2D fitter, TS map maker,
stereo TS finder, $z_{0}$ fitter, and output packer.\label{fig:3DTDesign}}
\end{figure*}

The 3DT extracts track parameters (sign of electric charge, $p_{T}$,
$\phi_{i}$, $\theta$, and $z_{0}$) from the input data which consists
of TSF, ETF, and 2DF data that changes every 32 ns. The TSF data has
four components which are the TS identification number (TS ID), left
right identification number (LR), priority layer identification number
(PR), and ``raw'' TDC which is relative to the revolution of the
beam. The ETF data has the event time relative to the revolution of
the beam. The 2DF data has five axial TS information related to a
found track.

The 3DT consists of multiple modules where the main modules are the
TS map maker, stereo TS finder and $z_{0}$ fitter which can be seen
in Fig. \ref{fig:3DTDesign}. The TS map maker integrates the TSF
data over a certain amount of time and creates an array of data which
we call TS map. The stereo TS finder relates stereo TSs with the tracks
that the 2DF finds. The $z_{0}$ fitter transforms the stereo TS data
into a geometrical representation and applies linear regression to
fit $z_{0}$ and $\theta$.

\subsection{TS map maker}

\begin{figure}
\begin{centering}
\includegraphics[width=1\columnwidth]{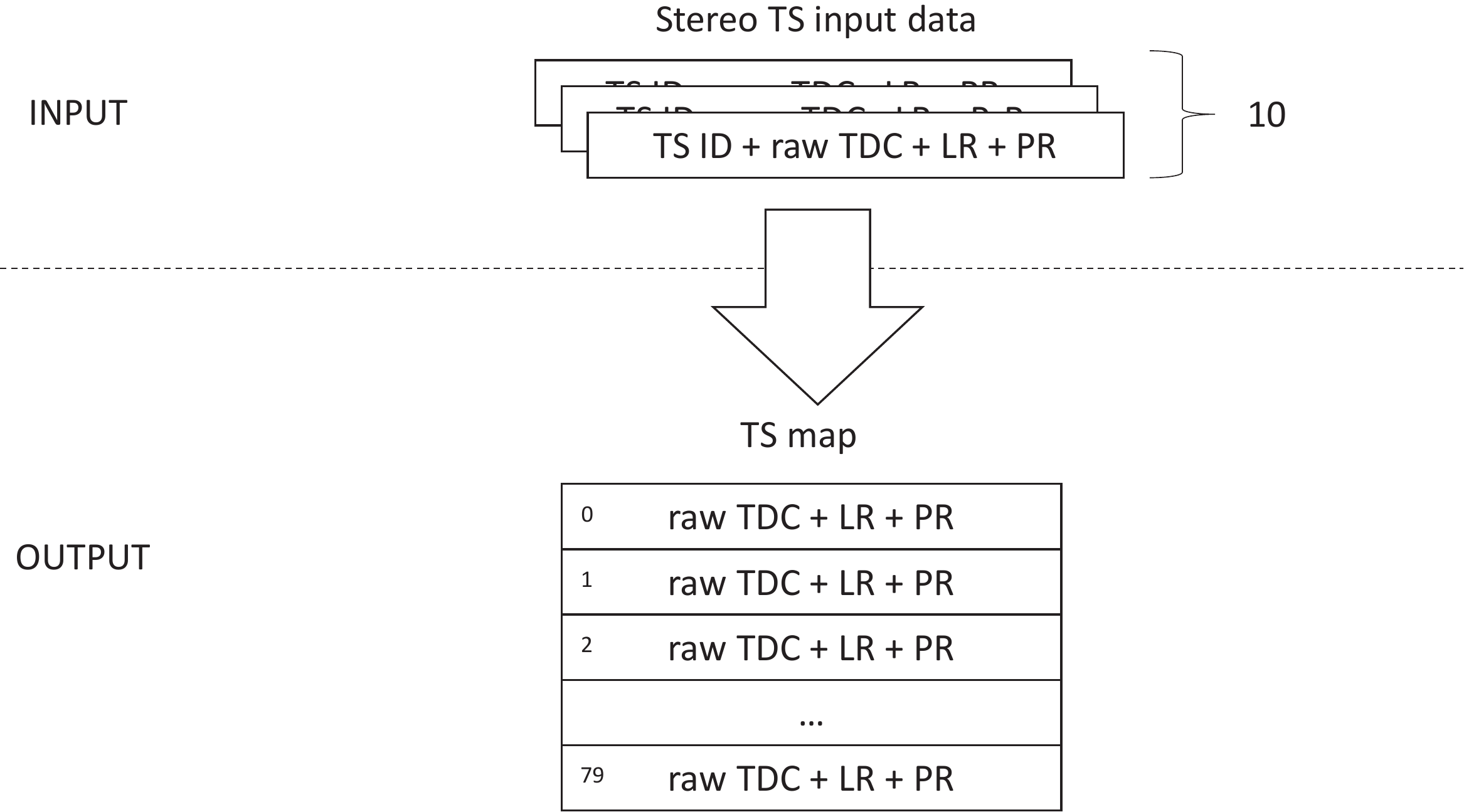}
\par\end{centering}
\caption{Logic flow of the TS map maker. This figure shows the structure of
the TS input and TS map. 10 TS information arrive at the 3DT every
32 ns. The input is transformed to an array of TS information arranged
according to the TS ID. \label{fig:TSMapMakerLogic}}

\end{figure}

\begin{figure}
\begin{centering}
\includegraphics[width=1\columnwidth]{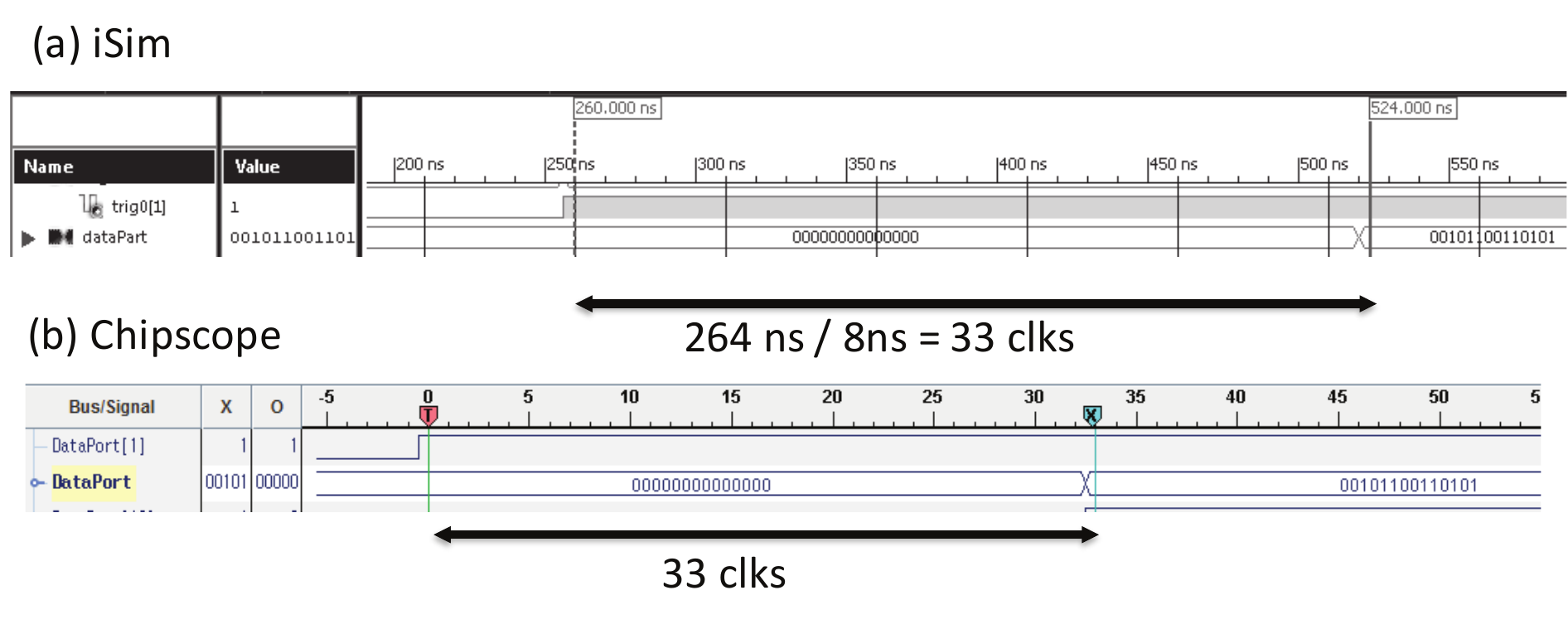}
\par\end{centering}
\caption{Comparison between ISim (a) and firmware test bench indicated as Chipscope
(b). The figure shows that the ISim results and timing exactly match
the firmware which was recorded using Chipscope. \label{fig:TSMapMakerTest}}
\end{figure}

Since the drift time for the CDC can be over than 500 ns, the TSF
data that changes every 32 ns should be stored and combined for a
certain amount of time. The TSF data consists of 10 TS information
for each 32 ns interval. This information is transformed to an array
of TS information where the index is the TS ID which we call TS map.
The structure of the input TSs and output TS map can be seen in Fig.
\ref{fig:TSMapMakerLogic}. The TS map is then sent to a module which
stores and combines the TS information. When new TS information for
a certain index enters the module, a counter which counts how long
the TS information has been saved is reset. When the counter reaches
a certain amount of time, the TS information is erased. 

Four stereo TSF data which cover half of the CDC enter the 3DT, so
that there are four of these TS map makers. Each TSF has a different
number of possible TS IDs, which are 80, 112, 144, and 176, so that
the number of indices of the TS maps follow these numbers. 

A VHDL \cite{VHDL} testbench was developed to test the developed
firmware for the TS map maker. It was shown that the TS map maker
works well as can be seen in Fig. \ref{fig:TSMapMakerTest}. The ISim
\cite{ISim} results and firmware results recorded with Chipscope
\cite{Chipscope} were compared. The timing and values of the results
were found to be the same.

\subsection{Stereo TS finder}

\begin{figure}
\begin{centering}
\includegraphics[width=1\columnwidth]{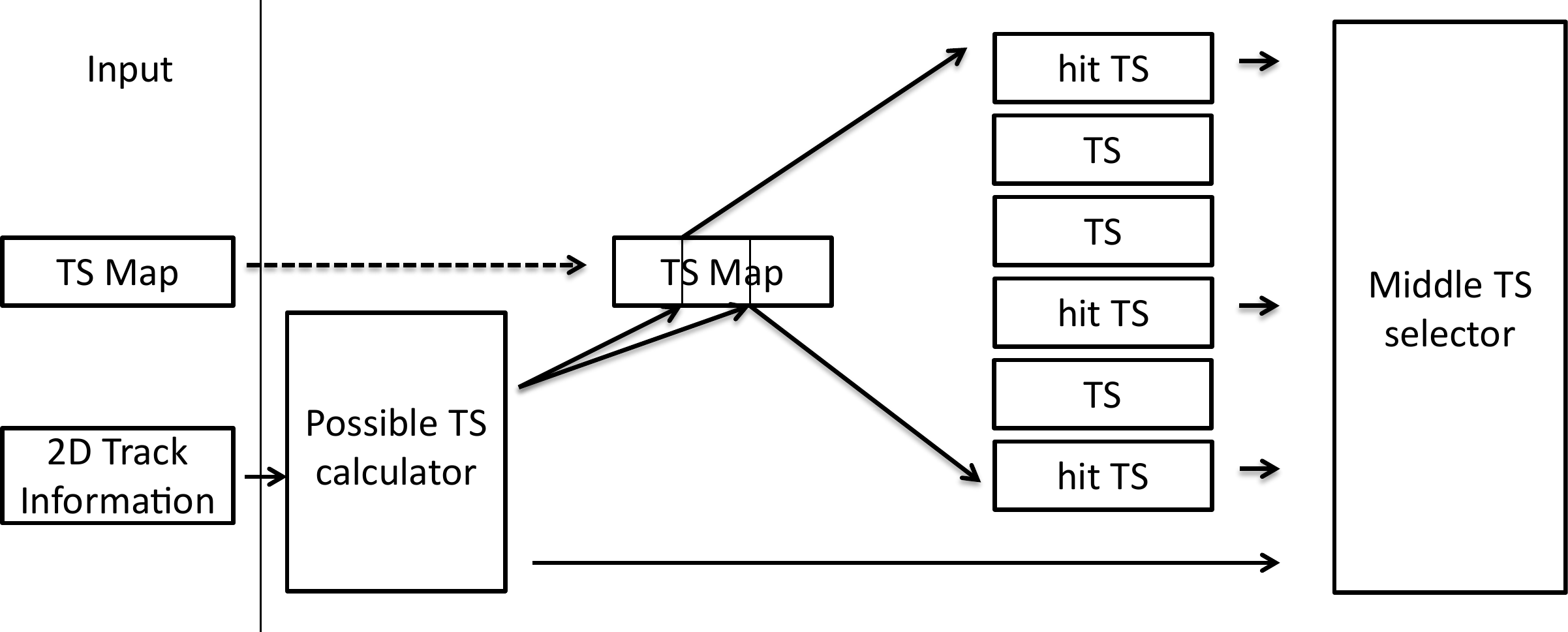}
\par\end{centering}
\caption{Logic flow of the stereo TS finder. This figure shows the detail of
how the related TS to a 2D track is found. Using the 2D information
the possible TSs are calculated. The TSs within the possible TSs are
selected. The TS which is hit and is near the middle is chosen to
be the related stereo TS.\label{fig:StereoTSFinderLogic}}

\end{figure}

\begin{figure}
\begin{centering}
\includegraphics[width=1\columnwidth]{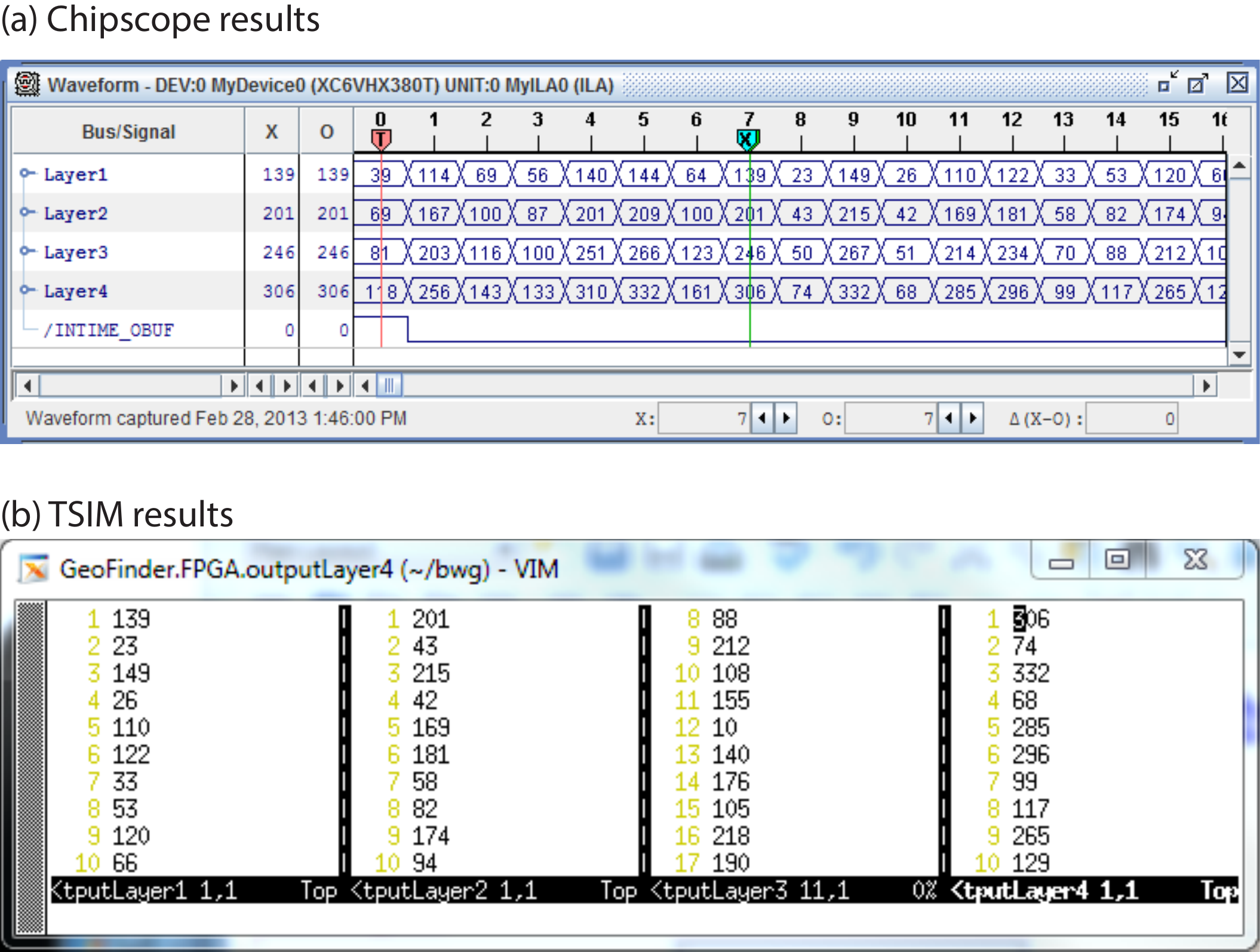}
\par\end{centering}
\caption{Comparison between firmware test bench indicated as Chipscope (a)
and TSIM (b). The figures show the chosen related stereo TS for each
stereo SL. The TSIM results and firmware results match perfectly.
\label{fig:StereoTSFinderTest}}
\end{figure}

After a track is found by the 2DF, related stereo TSs should be found.
By using the two dimensional track information, the geometrically
possible stereo TSs are calculated where a framework that automatically
generates VHDL was used for development \cite{VHDLAutomatic}. Between
the hit TSs for the possible TSs, the TS which is near the middle
is selected. The flow of the logic can be seen in Fig. \ref{fig:StereoTSFinderLogic}.

To calculate the geometrically possible stereo TSs, the location where
the track would pass if the stereo SL was an axial SL is calculated.
This location is calculated using the following equation according
to the sign of the electric charge,
\[
\phi_{\text{ax}}=\pm\arccos\left(\frac{r\rho}{2}\right)+\phi_{i}\mp\pi,
\]
where the radius of the CDC wire layer is $r$ and the curvature and
incident angle of the found track are $\rho$ and $\phi_{i}$. This
calculated phi location is transformed to the according TS ID by multiplying
a constant which depends on the SL. Since the geometrically possible
stereo TSs should be within ten TSs from the ID calculated above,
an array of 10 TSs is made. Between the hit TSs in the array, the
TS which corresponds to the closest point to the IP is selected.

{} The firmware was tested with a VHDL testbench where data was generated
from trigger simulation (TSIM). The results between TSIM and the firmware
output recorded with Chipscope were found to be exactly the same for
the stereo TS finder in each SL which can be seen in Fig. \ref{fig:StereoTSFinderTest}. 

\subsection{$z_{0}$ fitter}

\begin{figure}
\begin{centering}
\includegraphics[width=1\columnwidth]{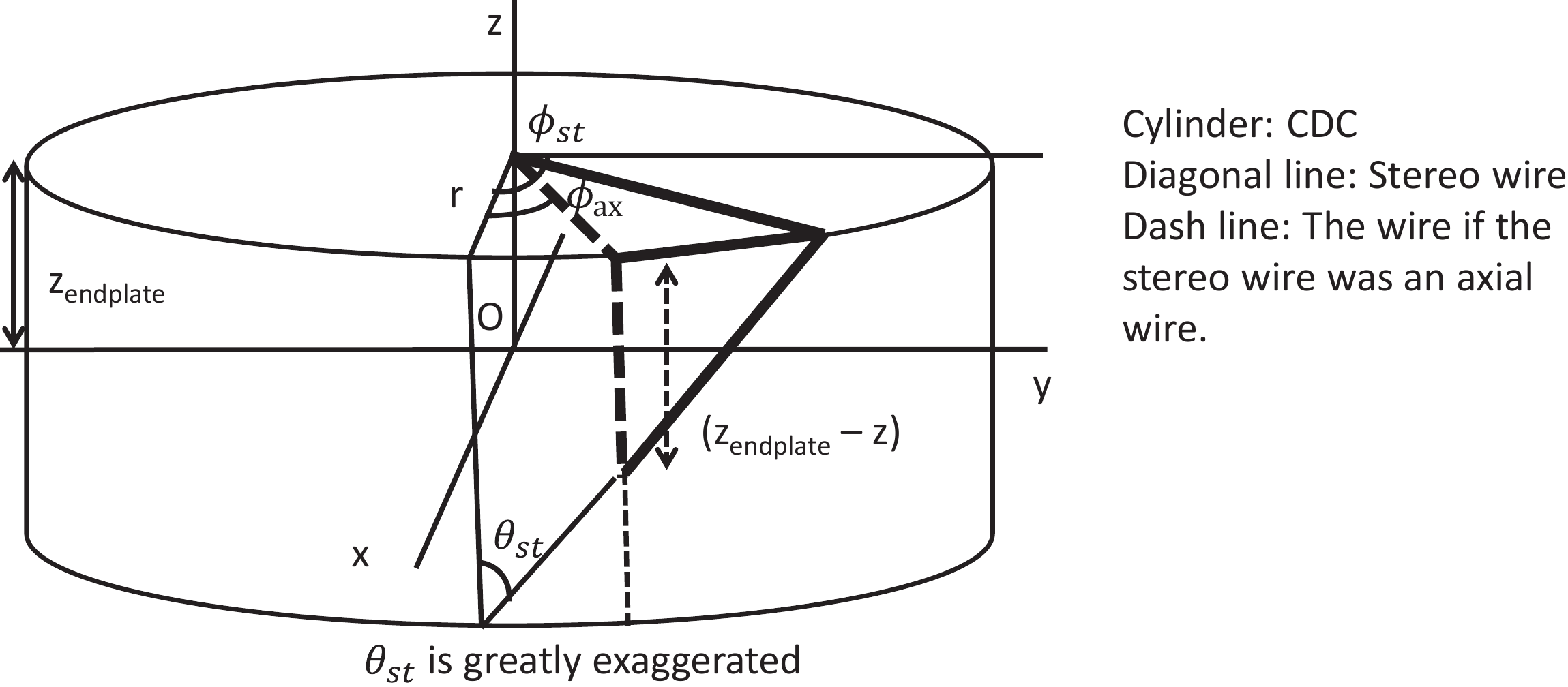}
\par\end{centering}
\caption{Geometry for calculating $z$. $z_{\text{endplate}}$ is the distance
from the IP to the end-plate. $r$ is the radius of the hit stereo
wire. $\phi_{\text{st}}$ is the phi of the hit stereo wire. $\phi_{\text{ax}}$
is the phi of the hit stereo wire if it was an axial wire. $\theta_{\text{st}}$
is the stereo angle of the stereo wire layer. The figure shows that
the length of $\left(z_{\text{endplate}}-z\right)\tan\theta_{st}$
is equal to $2r\sin\left(\frac{\phi_{\text{st}}-\phi_{\text{ax}}}{2}\right)$.
\label{fig:geometricZ}}
\end{figure}

\begin{figure}
\begin{centering}
\includegraphics[width=1\columnwidth]{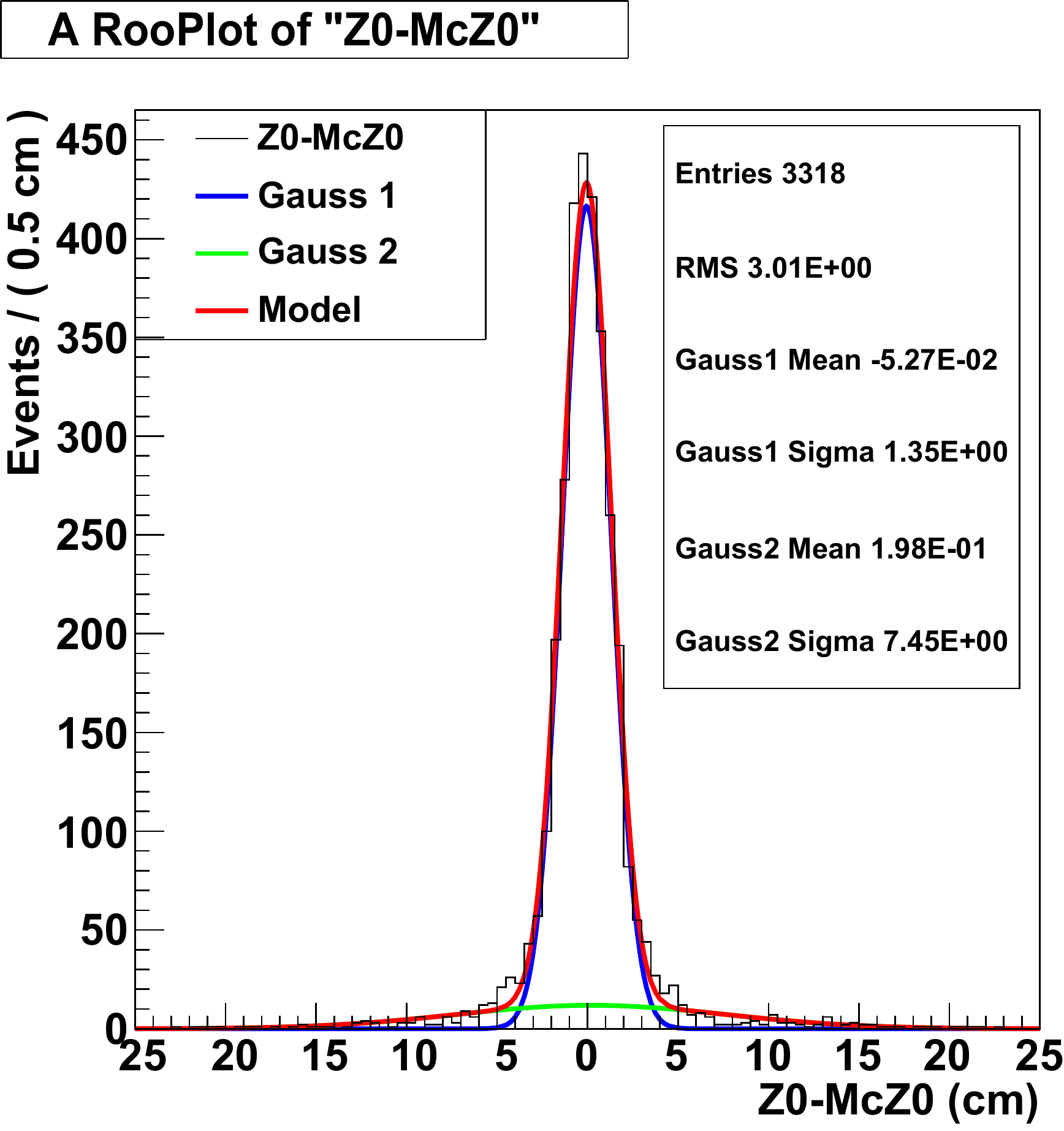}
\par\end{centering}
\caption{Resolution of $z_{0}$ using TSIM. A histogram of Monte Carlo truth
$z_{0}$ (MC) subtracted by the fitted $z_{0}$ can be seen. The core
peak was fitted with a Gaussian. The fitted Gaussian's sigma is 1.35
cm. \label{fig:z0FitterResolution}}

\end{figure}

The related stereo TS information is transformed to a geometric representation
and then fitted to get track parameters. We first find TDC for each
stereo TS by subtracting the ``raw'' TDC of each stereo TS with
``raw'' event time. The TDC is converted to a length dimension which
we call drift length using a look-up table that has the x-t curve
of the CDC. Using the LR information from the TSF, the drift length
is added or subtracted to the TS phi position which we call fine phi
position. Using two dimensional track parameters and related stereo
TS fine phi positions, the $z$ location for each stereo TS is calculated
with the following equation,
\[
z=\frac{z_{\text{endplate}}-2r\sin(\frac{\phi_{\text{st}}-\phi_{\text{ax}}}{2})}{\tan\theta_{\text{st}}},
\]
where $z_{\text{endplate}}$ is the distance from the IP to the end
plate of the CDC, $\theta_{\text{st}}$ is the angle of the stereo
wires, and $\phi_{\text{st}}$ is the fine phi position the stereo
TS. The geometric representation of this equation can be seen in Fig.
\ref{fig:geometricZ}. The arc length of the track in the two dimensional
($x,y$) plane for the radius of each stereo SL is calculated with
the following equation,
\[
s=\arcsin\left(\frac{r\rho}{2}\right),
\]
where $s$ is the arc length. The track of charge particles in an
uniform magnetic field is a helix, so that the two dimensional arc
length and $z$ of stereo TSs should have a linear relationship. We
construct a $\chi^{2}$ as below to fit the $z_{0}$ and $\cot\theta$
using the geometric representation of the stereo TSs,
\[
\chi^{2}=\sum\limits _{i}\frac{[z_{i}-(\cot\theta\times s_{i}+z_{0})]^{2}}{\sigma_{i}^{2}},
\]
where $\sigma_{i}$ is the error of $z_{i}$. The $\cot\theta$ and
$z_{0}$ which minimize the $\chi^{2}$ can be found analytically
as below 
\[
\left[\begin{array}{c}
\cot\theta\\
z_{0}
\end{array}\right]=\left[\begin{array}{c}
\frac{(\sum\limits _{i}\frac{1}{\sigma_{i}^{2}})(\sum\limits _{i}\frac{s_{i}z_{i}}{\sigma_{i}^{2}})-(\sum\limits _{i}\frac{s_{i}}{\sigma_{i}^{2}})(\sum\limits _{i}\frac{z_{i}}{\sigma_{i}^{2}})}{(\sum\limits _{i}\frac{1}{\sigma_{i}^{2}})(\sum\limits _{i}\frac{s_{i}^{2}}{\sigma_{i}^{2}})-(\sum\limits _{i}\frac{s_{i}}{\sigma_{i}^{2}})(\sum\limits _{i}\frac{s_{i}}{\sigma_{i}^{2}})}\\
\frac{-(\sum\limits _{i}\frac{s_{i}}{\sigma_{i}^{2}})(\sum\limits _{i}\frac{s_{i}z_{i}}{\sigma_{i}^{2}})+(\sum\limits _{i}\frac{s_{i}^{2}}{\sigma_{i}^{2}})(\sum\limits _{i}\frac{z_{i}}{\sigma_{i}^{2}})}{(\sum\limits _{i}\frac{1}{\sigma_{i}^{2}})(\sum\limits _{i}\frac{s_{i}^{2}}{\sigma_{i}^{2}})-(\sum\limits _{i}\frac{s_{i}}{\sigma_{i}^{2}})(\sum\limits _{i}\frac{s_{i}}{\sigma_{i}^{2}})}
\end{array}\right].
\]
The algorithm was developed using a framework that automatically generates
VHDL and validated using TSIM. The resolution of $z_{0}$ was found
to be 1.35 cm which can be seen in Fig. \ref{fig:z0FitterResolution}.
A VHDL testbench was developed for the firmware. The firmware output
and TSIM were found to be exactly the same \cite{VHDLAutomatic}.

\section{CONCLUSIONS}

We have developed firmware that extracts track parameters from the
input of the 3DT. The logic we developed that stores the input, finds
related TS to a track, transforms the stereo TS to geometric representations,
and does linear regression has been shown. Using the extracted track
parameters, beam induced backgrounds which do not originate near the
IP can be effectively rejected. 

\section*{Acknowledgments}

J. B. acknowledges support from the National Research Foundation of
Korean Grants No. 2016K1A3A7A09005587. We thank Korea Institute of
Science and Technology Information network group for KREONET network
support.

\end{document}